\documentclass[useAMS,usenatbib]{mn2e}

\usepackage{graphicx}
\usepackage{amsmath}
\usepackage{subfigure}
\title[Revisiting the parametrization of Equation of State of Dark Energy via SNIa Data]{Revisiting the parametrization of Equation of State of Dark Energy via SNIa Data}

\author[Dao-Jun Liu, Xin-Zhou Li, Jiangang Hao and Xing-Hua Jin]{Dao-Jun Liu$^{1}$\thanks{E-mail:
djliu@shnu.edu.cn (DJL)} Xin-Zhou Li$^{1}$\thanks{E-mail:
kychz@shnu.edu.cn (XZL)} Jiangang Hao$^{2}$\thanks{E-mail:
jghao@umich.edu (JGH)} and Xing-Hua Jin$^{3}$\thanks{E-mail:
cufa1@shnu.edu.cn (XHJ)}\\
$^{1}$Center for Astrophysics, Shanghai Normal University,
100 Guilin Road, Shanghai, 200234, China\\
$^{2}$Department of
physics, University of Michigan, 500 E. University Ave., Ann Arbor, MI 48109, USA\\
$^{3}$Cambridge International Centre of Shanghai Normal University,
120 Guilin Road, Shanghai, 200234, China}

\begin{document}

\date{  Received ????.??.??; Accepted ????.??.?? }

\pagerange{\pageref{firstpage}--\pageref{lastpage}} \pubyear{}

\maketitle

\label{firstpage}

\begin{abstract}
In this paper, we revisit the parameterizations of the equation
of state of dark energy and point out that comparing merely the
$\chi^2$ of different fittings may not be optimal for choosing the
"best" parametrization. Another figure of merit for evaluating different parametrizations based
on the area of the $w(z) - z$ band is proposed. In light of the analysis of some  two-parameter parameterizations and models based on available SNIa data,  the area of $w(z)-z$ band seems to be a good figure of merit, especially in the situation that the value of $\chi^2_{\rm min}$ for different parametrizations are very close. Therefore, we argue that both the area of the $w(z)-z$ band and $\chi^2_{\rm min}$  should be synthetically considered for choosing a better parametrization of dark energy in the future experiments.
\end{abstract}


\section{introduction}

Current observations, such as those of CMB anisotropy
\citep{WMAP},
Supernovae type Ia \citep{gold,SNIa,Davis07}
 and large scale structure \citep{SDSS,2dF}, converge on the fact that a spatially homogeneous
and gravitationally repulsive energy component, referred as dark
energy, account for about 70\% of the energy density of Universe. Some heuristic models that roughly describe the observable consequences of dark energy were suggested in recent years, a number of them stemming form fundamental physics and other being purely phenomenological.
However, the nature of dark energy still remain mysterious to
physicists and astronomers although many possible candidates have
been proposed. Dark energy present in the equations of cosmological dynamics through its effective energy density and pressure. The ratio of pressure to energy density (the equation of state) is very important in the Friedmann equation regardless of its physical origin. If dark energy is some kind of dynamical fluid and its equation of state would likely not be constant, but would vary with redshift $z$ or equivalently with cosmic time. The impact of dark energy (whether dynamical or a constant) on cosmological observations can be expressed in term of $w(z)=p(z)/\rho(z)$ which is to be measured through either the cosmic expansion history $H(z)$(obtained, for example, using supernova data) or through large-scale structure. Therefore, it is sagacious to study the parameterization of the equation of state of dark energy empirically with as few prior assumptions as possible.

To reveal the nature of dark energy and narrow
down the candidate list, a very powerful measure is to map out the
evolution of the equation of state as redshift changes.
However, in data fitting, we need to parameterize the equation of
state $w(z)$ in simple form and then constrain the evolution of
$w(z)$ in terms of the parameters we introduced in our
parametrization except the case in which $w(z)$ is already
such as in the quintessence field \citep{a1,a2,a3,a4,a5}, phantom field \citep{b1,b2,b3,b4}, or Chaplygin gas model \citep{c1,c2}.  Unquestionably, the way we parameterize
the equation of state is bound to affect our ability to extract information
from the data. There are many different parameterizations have been
introduced based on simplicity and the requirement of regular
asymptotic behaviors \citep{p1,p2,p3}. However, will these choice of
parameterizations give us maximum power to extract information from
the data? Some analysis existing in literatures compared the
different parameterizations by looking at their corresponding
$\chi^2$, which are justified by the generalized likelihood ratio test in
statistics. But this measure is no longer fair when the $\chi^2$ is
small but the curvature of the likelihood function is very big, meaning that
the constraints on the parameters are loose although the resulting
$\chi^2$ is small.

In this paper, we introduce another figure of merit in analogous
to Ref.\citep{DETF1,DETF2}, the area of the $w(z)-z$ band to
evaluate the performance of different parameterizations. The
justification of this measure lies in that our ultimate goal is to
constrain the shape of $w(z)$ as much as we can from the data.
In our analysis, we will compare the parameterizations with identical
number of parameters. Note that comparing parameterizations with different
number of parameters based on Akaike's Information Criterion (AIC), Bayesian Information Criterion (BIC) or other criteria is
arbitrary in the sense of the criteria one chooses. In a sensible Bayesian method, a model is penalized for having a larger number of parameters that gives a reasonable fit (not the best fit compared to models with more parameters) is awarded with increased evidence for that model, (see for example, Ref.\citep{Liddle06}). However, this is not what we are concerned and the purpose of this paper is just to show what is the best way to parameterize the equation of
state of dark energy for a variety of prevalent models with \emph{identical} number of parameters. Our results
show that the widely used parameterization, $w(z)=w_0+w_1z/(1+z)$, is not the one that
can tell us most of the information of $w(z)$ in two-parameter
parameterization family based on the SNIa data.

Among the many observations that can help to constrain the shape
of $w(z)$, SNIa data provide most sensitive and straightforward
constraints. Therefore, in this paper, we will study the
effects of different parameterizations on our understanding of the
evolution of dark energy based on SNIa data.

The outline of the paper is as follows. In section II, we discuss the expansion history of the universe and the observational variables from the supernova experiments. In section III, two parameterization families and some prevalent models of equation of state of dark energy is introduced. Throughout the paper, we only consider two-parameter models and parameterizations. In section IV, the method and results of the analysis is presented.  In the last, we conclude with some remarks on the choice of parameterization of dark energy.

\section{the expansion history of the universe and supernova  }
In the framework of standard cosmological model, assuming a spatially flat ($k = 0$) Friedmann universe, the equations
governing the expansion of the universe are
\begin{equation}\label{H}
    H^2=H_0^2\left[\Omega_M(1+z)^3+(1-\Omega_M)f(z)\right],
\end{equation}
and
\begin{equation}\label{q}
    q=\frac{3w(z)(1-\Omega_M)+1}{2},
\end{equation}
where $H\equiv \dot{a}/a$ is the Hubble parameter, $q\equiv -\ddot{a}/aH^2$ is the deceleration parameter, $\Omega_M\equiv \rho_M/\rho_C$ is the cosmic matter density parameter and $w(z)$ is the dark energy equation of state, which is defined by
\begin{equation}
w\equiv \frac{p_{DE}}{\rho_{DE}}.
\end{equation}
The dark energy density parameter $\rho_{DE}$ evolves as $\rho_{DE}(z)=\rho_{DE}^0f(z)$ and
\begin{equation}\label{f(z)}
    f(z)=\exp\left[3\int_0^z\frac{1+w(z')}{1+z'}dz'\right].
\end{equation}
To date, Supernovae Type Ia (SNIa) provide the most direct indication of the accelerating expansion of the universe. For the distant SNIa, one can directly observe their apparent magnitude $m$ and redshift $z$, because the absolute magnitude $M$ of them are assumed to be constant, i.e., SNIa are standard candles. The luminosity distance $d_L(z)$ is the "meeting point" between the observed $m(z)$ and theoretical prediction $H(z)$:
\begin{equation}
m(z)=M+5\log_{10}\left[\frac{d_L(z)}{\mathrm{Mpc}}\right]+25
\end{equation}
and
\begin{equation}
d_L(z)=(1+z)\int_0^z\frac{c\,dz'}{H(z')}.
\end{equation}

\section{parameterizations of dark energy}\label{parameterizations}

Although $H(z)$ is more directly related to the observable luminosity distance and then is easier to measure more accurately, in order to investigate the evolution of dark energy with time and the scale factor, constraints on $w(z)$  is essentially equivalent to that of $H(z)$ and is also crucial for understanding the nature of dark energy \citep{Huterer}. Since $w(z)$ is a continuous function with an infinite number of values at a finite redshift range, $w(z)$ must be modeled using just a few parameters whose values are determined by fitting to observations. A merit of using $w(z)$ with a particular parametrization is to compare the performance of different experiments.  Note here that no single parameterization can represent all possibilities for $w(z)$. A reasonable parameterization must be accorded with the demand that dark energy is important at late times and insignificant at early times.

There exist plenty of parameterizations for the equation of state $w(a)$ \citep{p1,p2,p3} where $a=(1+z)^{-1}$, but most of them are purely phenomenological. Maybe, we should consider some of them in the sense that they are generalized from the behavior of physically motivated sets of models \citep{Linder07}.

For single parameter models, e.g. $w=costant$, no dynamics is embodied and can not parameterize the rate of change of $w$ and then high fine-tuning is needed.  The physical symmetry motivated one-parameter models, such as topological defects, are not consistent with the observation data. More parameters mean more degrees of freedom for adaptability to
observations, at the same time more degeneracies in the determination of parameters. For models with more than two parameters, they lack predictability and even the next generation of experiments will not be able to constrain stringently \citep{LH05}. Therefore, we only consider the two-parameter models in this paper. Of course, two-parameter models also have  limitations; for example, it is hard to describe rapid variation of $w(z)$ in most of these models.

Various two-parameter parameterization approaches have been proposed in the literatures. The simplest way to parameterize the rate of change of $w$ is to write the first-order Taylor expansion. This is the linear-redshift parameterization (Linear)\citep{Huterer01,Hut02}, which is given by
\begin{equation}
w=w_0+w_1z.
\end{equation}
This parameterization is not viable as
it diverges for $z >> 1$ and therefore incompatible with the constraints from CMB \citep{CMBconstaint} and BBN \citep{BBNC}. The Upadhye-Ishak-Steinhardt parameterization (UIS) \citep{UIS} can avoid above problem,
\begin{equation}
w=\begin{cases}
w_0+w_1z, & \text{if $z<1$},\\
w_0+w_1, & \text{if $z\geq1$}.
\end{cases}
\end{equation}

We here mainly consider the following two commonly used two-parameter parameterization families:\\
\underline{Family I}:
  \begin{equation}\label{n}
    w=w_0+w_1\left(\frac{z}{1+z}\right)^n,
\end{equation}
\underline{Family II}:
\begin{equation}\label{dn}
w=w_0+w_1\frac{z}{(1+z)^n},
\end{equation}
where $w_0$ and $w_1$ are two undecided parameters, $n=1,2,\cdots$. Both of the parameterization families have the reasonable asymptotical behavior at high redshifts. The case with $n=1$ in the above parameterization approaches is the same as the most popular parameterization introduced by Chevallier and Polarski \citep{CPL1} and Linder \citep{CPL2} parameterization (CPL). This simple
parameterization is most useful if dark energy is important at late times and
insignificant at early times. The one with $n=2$ in Family II is the Jassal-Bagla-Padmanabhan parameterization (JBP), which can model a dark energy component that has the same equation
of state at the present epoch and at high redshifts, with rapid variation
at low $z$\citep{JBP}.

Another two-parameter parameterization of dark energy equation of state, we consider here, comes from the direct $H(z)$ parameterization, first suggested by Sahni \textit{et al.}~\citep{P2},
\begin{eqnarray}
 H^2&=&H_0^2\left[\Omega_M(1+z)^3+\Omega_2(1+z)^2\right.\nonumber\\
 &+&\left.\Omega_1(1+z)+(1-\Omega_M-\Omega_1-\Omega_2)\right],
\end{eqnarray}
which is corresponding to an effective equation of state of dark energy (P2)
\begin{equation}
w(z)=-1+\frac{(1+z)[\Omega_1+2\Omega_2(1+z)]}{3\left[\Omega_2z^2+(\Omega_1+2\Omega_2)z+1-\Omega_M\right]}.
\end{equation}
On the other hand, there are also two-parameter models that have direct physical meanings. For example, generalized Chiplygin gas model (GCG) \citep{GCG1,GCG2,GCG3,GCG4}, which has effective equation of state
\begin{equation}
w(z)=-1+\frac{a_2(1+z)^3(a_1-1)}{z(z^2+3z+3)a_1-(1+z)^3}.
\end{equation}

\section{Method and results}
We use the Fisher matrix methods to compute the covariance matrix for the parameters $w_i$. If the parameters $w_i$ gives the true underlying distribution $\bar{w}$, then a chi square distribution of data values is
in proportion to $\exp(-\chi^2/2)$, where $\chi^2$ is determined by
\begin{equation}
\chi^2=\sum_{i}\frac{(\mu_i-\mu^{th})^2}{\sigma_i^2},
\end{equation}
where $\sigma_i$ is error of the distance modulus $\mu_i$ and $\mu^{th}$ is theoretical prediction to the data. For SNIa data we use here, $\mu^{th}=\mu^{th}(z;\{w_i\})=m(z)-M$. Using Bayes' theorem with uniform prior to the parameter, the likelihood of a parameter estimate can be described as a Gaussian with the same $\chi^2$, which is now viewed as a function of parameters $\chi^2=\chi^2(\{w_i\})$.  The distribution of errors in the measured parameters is in the limit of high statistics proportional to (see, for example, Ref.\citep{DETF1})
$$\exp\left(-\frac{1}{2}F_{ij}\sigma_{w_i} \sigma_{ w_j}\right),$$ where the Fisher matrix $F_{ij}$ is defined by
\begin{equation}
F_{ij}=\frac{1}{2}\left<\frac{\partial^2\chi^2}{\partial w_i \partial w_j}\right>,
\end{equation}
and $< \cdots >$ means average over realizations of the data. The covariance matrix of the parameters is simply the inverse of the Fisher matrix,
\begin{equation}
C_{ij}=(F^{-1})_{ij}.
\end{equation}
The error on the equation of state $w(z)$ is given by \citep{Nesseris}
\begin{equation}
\sigma_w^2=\sum_{i=1}^{N}\left(\frac{\partial w}{\partial w_i}\right)^2C_{ii}+2\sum_{i=1}^{N}\sum_{j=i+1}^{N}\frac{\partial w}{\partial w_i}\frac{\partial w}{\partial w_j}C_{ij},
\end{equation}
where $N$ is the number of the free parameters. $\sigma_w$ is function of $z$, we define the area of $w(z)-z$ band as
\begin{equation}\label{s}
s=2\int_{z_l}^{z_h}\sigma_{w}(z)dz,
\end{equation}
where the integral interval $(z_l, z_h)$ is taken as $(0,z_{max})$.

We make use of the full Gold dataset \citep{gold} ($157$ data points, $0 < z < z_{max}=1.755$) and  the combined Essence, Hubble, SNLS and nearby supernovae catalog as compiled by \citep{Davis07}, for a total of 192 supernovae, respectively, assuming  a flat universe with energy density in matter $\Omega_M=0.3$. The value of the Hubble constant $H_0$ is marginalized analytically.

\begin{figure*}
\begin{minipage}{\textwidth}
\begin{center}
\mbox{\subfigure{\includegraphics[width=0.48\textwidth]{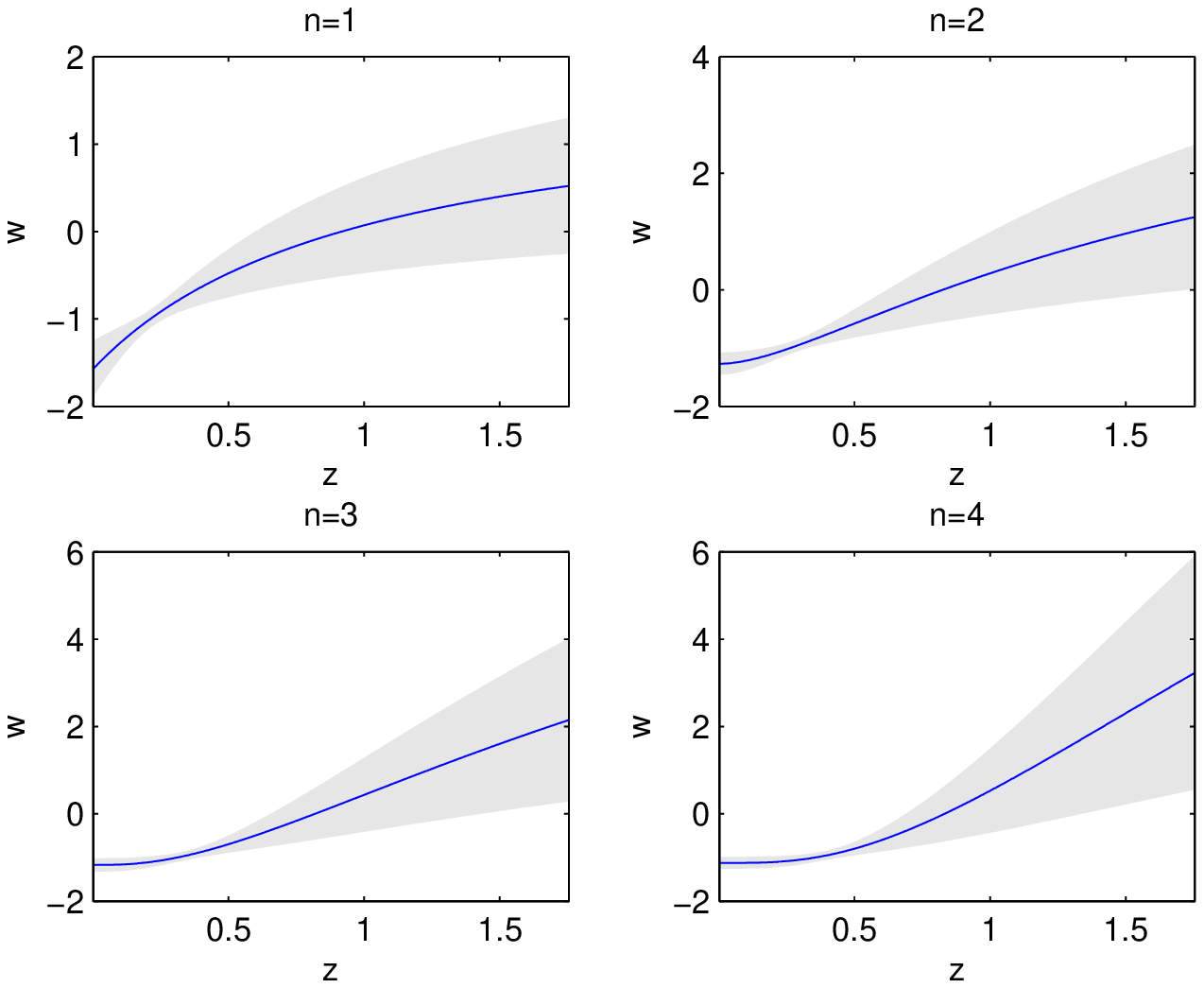}}
      \subfigure{\includegraphics[width=0.48\textwidth]{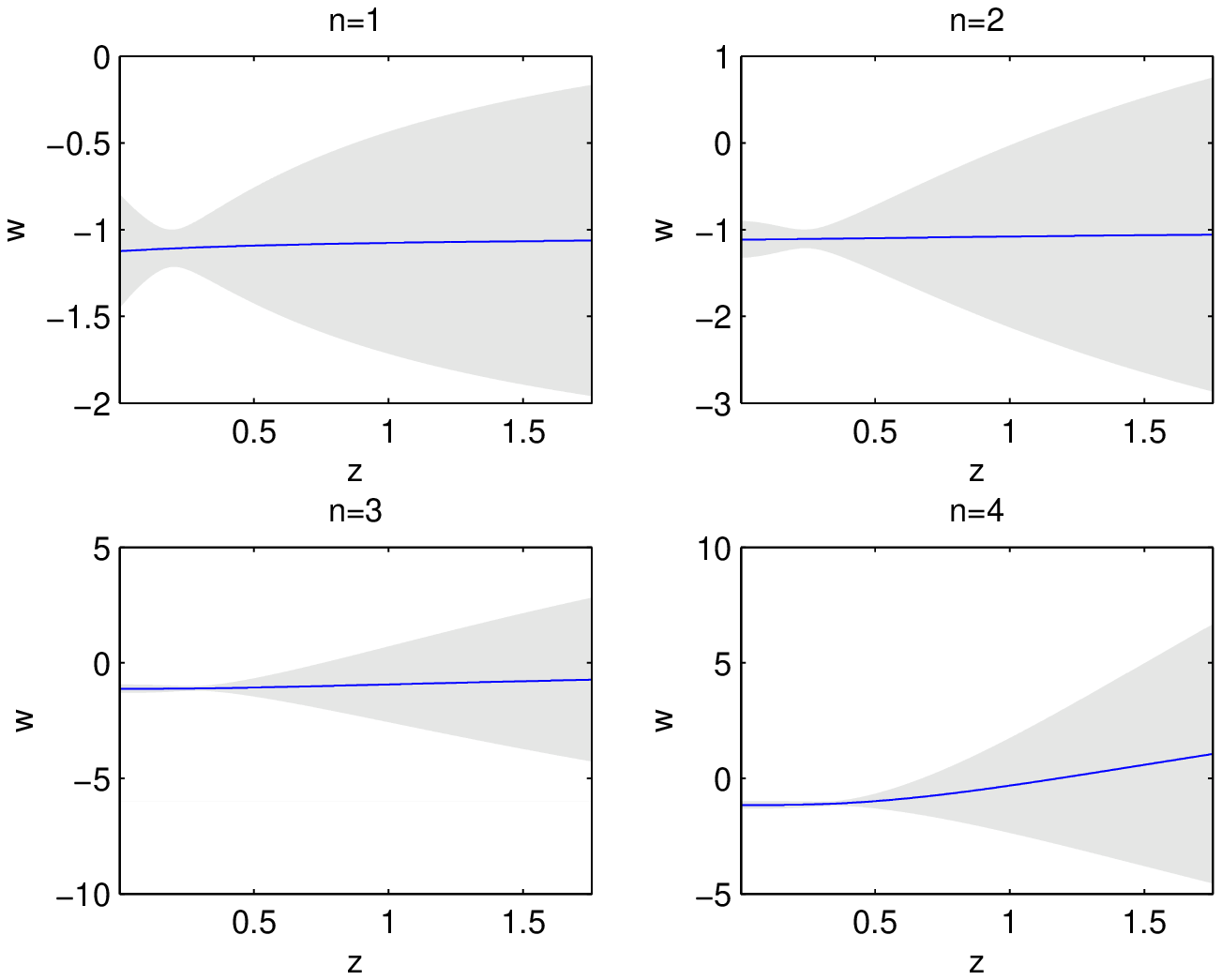}}}
\caption{$w(z)-z$ band for family I parametrizations. The left four panels are obtained by using 157 Gold data \citep{gold} and the right four are obtained by using latest 192 SNIa data \citep{Davis07,Wood07,SNIa}.}\label{fig1}
\end{center}
\end{minipage}
\end{figure*}

\begin{figure*}
\begin{minipage}{\textwidth}
\begin{center}
\mbox{\subfigure{\includegraphics[width=0.48\textwidth]{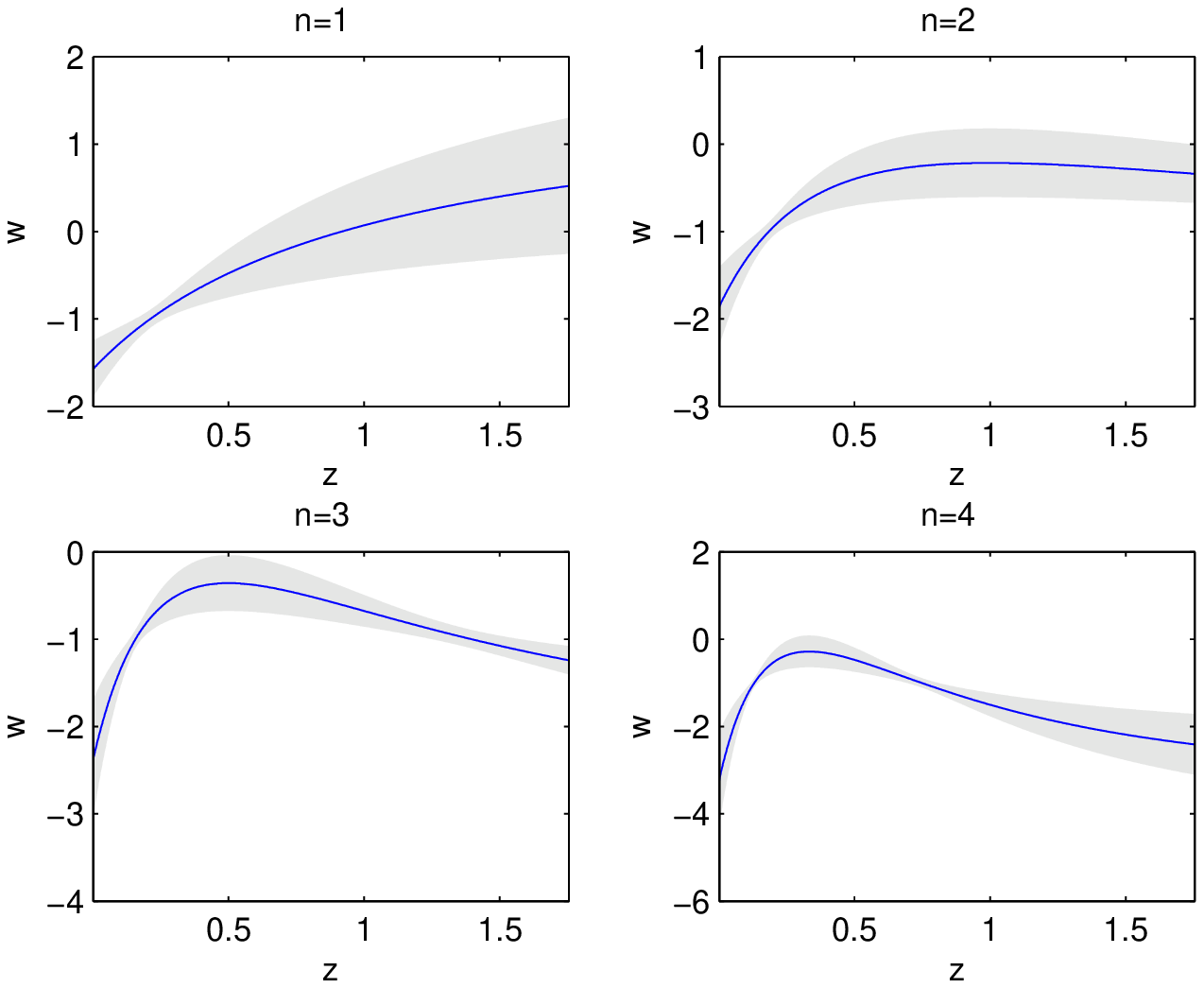}}
      \subfigure{\includegraphics[width=0.48\textwidth]{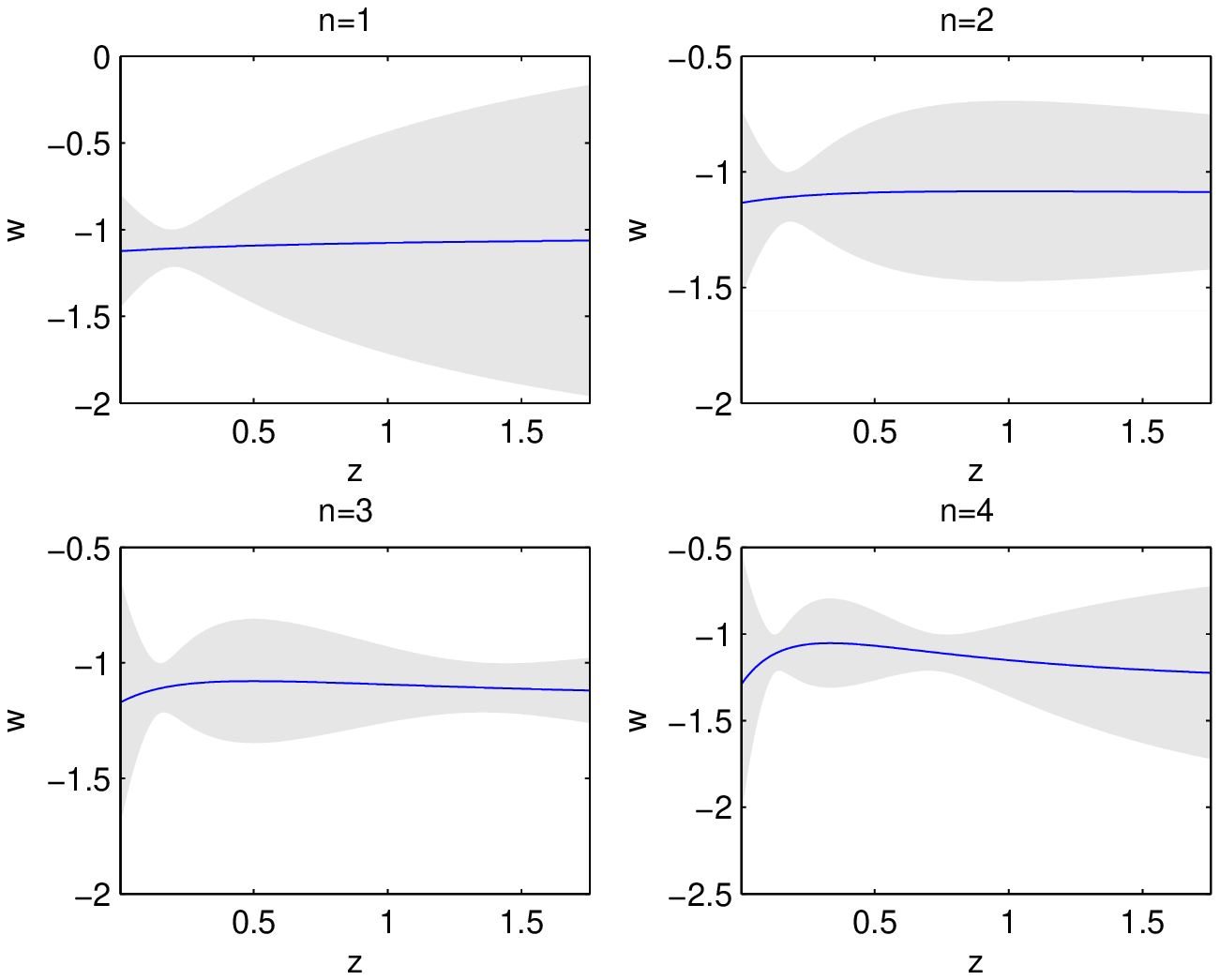}}}
\caption{$w(z)-z$ band for family II parametrizations. The left four panels are obtained by using 157 Gold data \citep{gold} and the right four are obtained by using latest 192 SNIa data \citep{Davis07,Wood07,SNIa}.}\label{fig2}
\end{center}
\end{minipage}
\end{figure*}

\begin{figure*}
\begin{minipage}{\textwidth}
\begin{center}
\mbox{\subfigure{\includegraphics[width=0.48\textwidth]{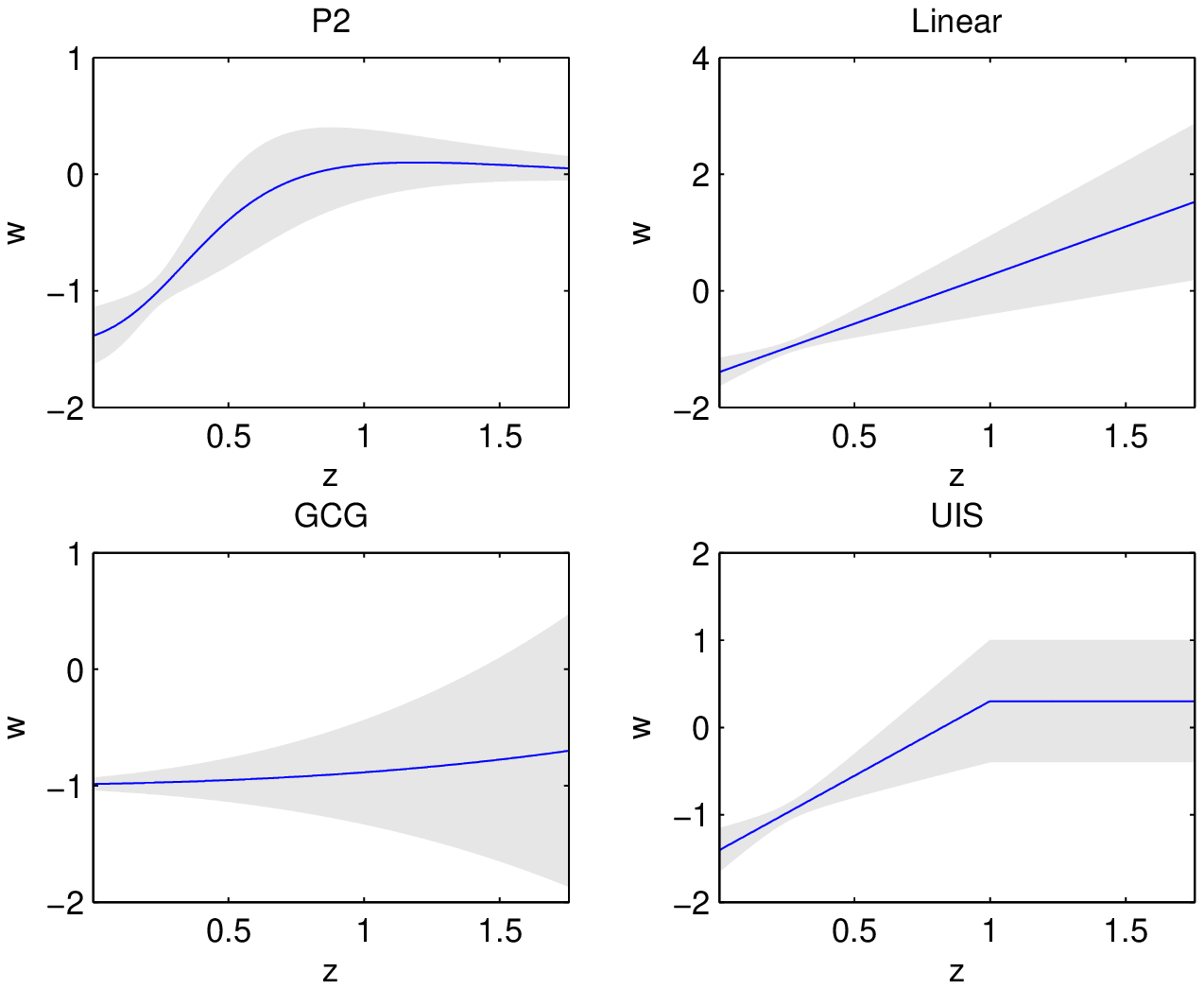}}
      \subfigure{\includegraphics[width=0.48\textwidth]{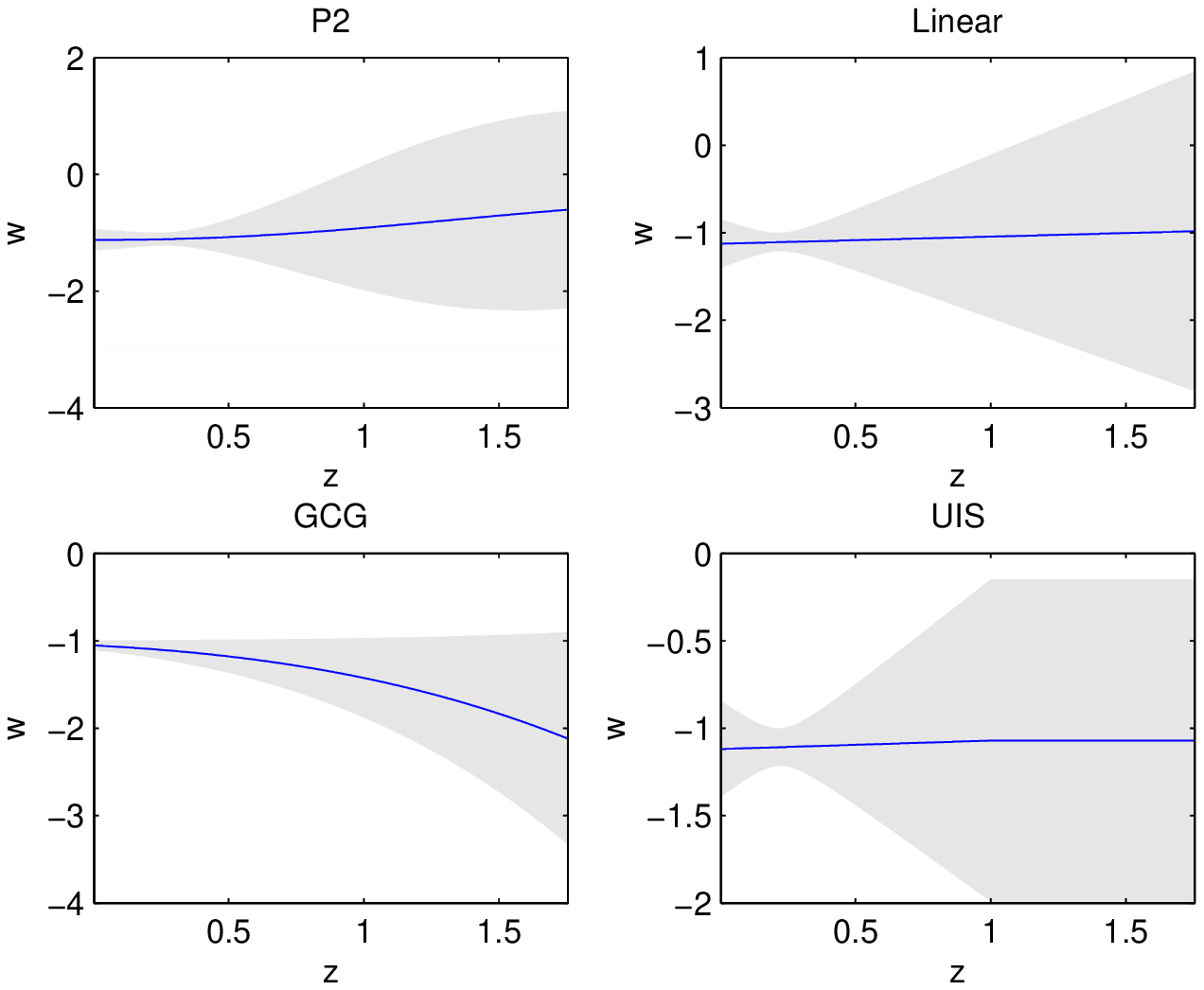}}}
\caption{$w(z)-z$ band for for some prevalent models. The left six panels are obtained by using 157 Gold data \citep{gold} and the right six are obtained by using latest 192 SNIa data \citep{Davis07,Wood07,SNIa}.}\label{fig3}
\end{center}
\end{minipage}
\end{figure*}

\begin{figure}
\begin{center}
\includegraphics[width=0.5\textwidth]{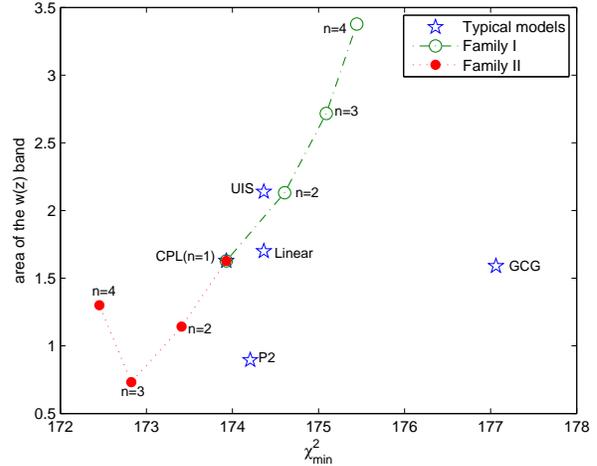}
\caption{ The portrait of the $\chi^2_{\rm min}-s$ phase of parametrizations of family I and family II and some prevalent models obtained by using 157 Gold SNIa data. }
\label{fig6}
\end{center}
\end{figure}

\begin{figure}
\begin{center}
\includegraphics[width=0.5\textwidth]{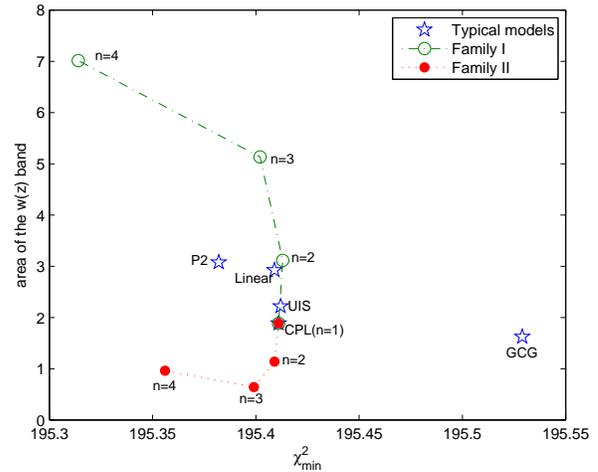}
\caption{ The portrait of the $\chi^2_{\rm min}-s$ phase of parametrizations of family I and family II and some prevalent models obtained by using newly compiled 192 SNIa data. }
\label{fig7}
\end{center}
\end{figure}

The main results are listed in Table \ref{table1} and Table \ref{table2}. Table \ref{table1} shows the minima of $\chi^2$, the best-fit values of the free model parameters and their standard deviations in the two parametrization families and several prevalent dark energy models introduced in section \ref{parameterizations}. The values of the corresponding areas of the $w(z)-z$ bands of different parameterizations and models are also shown in Table \ref{table1}. All of these quantities are worked out by using $157$ Gold SNIa data.  As a comparison, Table \ref{table2} shows the results of the same physical quantities  based on the newly compiled 192 SNIa data. To better explain the results, we plot the $w(z)-z$ bands for parameterizations of family I, family II and the selected prevalent models in figure \ref{fig1}, figure \ref{fig2} and figure \ref{fig3}, respectively. In figure \ref{fig6} and figure \ref{fig7}, we show the portraits of $\chi^2_{\rm min}-s$ phase of parameterizations of family I and family II and the selected prevalent models in the light of the results obtained by using 157 Gold SNIa data and newly compiled 192 SNIa data, respectively.

As is shown in figure \ref{fig6}, it is clear that for parameterizations in family I, the minima of $\chi^2_{\rm min}$ and $s$ are  coincide with each other at $n=1$, which corresponds to the widely used CPL parametrization. However, in family II, the minimum of $\chi^2_{\rm min}$ does not coincide with the one of $s$. The minimum of $\chi^2_{\rm min}$ is located at $n=4$, while the minimum of $s$ is located at $n=3$.
Note that $n=1$ parametrization (i.e. CPL) have neither the minimum $\chi^2_{\rm min}$ nor the minimum $s$. Therefore, if we have to choose a parametrization in family II, $n=1$ parametrization will not be preferred, and the two most competitive parametrizations are $n=3$ and $n=4$. But as the difference between two $\chi^2_{\rm min}$ is much less than the difference between two $s$, we would prefer the $n=3$ parameterization.  For the prevalent models we investigated here, the value of $\chi^2_{\rm min}$ of GCG model is much greater than those of CPL, UIS, Linear and P2 models, so it is not preferred in the sense of data fitting, but it may still be interesting because of its physical meaning.

As an improvement and extension of earlier data, the newly compiled data set \citep{Davis07} provide us more sample data.  Compare figure \ref{fig7} with figure \ref{fig6}, we find that although there exit minor changes between the results based on the 192 newly compiled SNIa data and those  based on the 157 gold data, the main results remain unchanged. Firstly,  for parametrizations in family I, the best one is  still $n=1$ due to its smallest area of $w(z)$ band, albeit the minimum value of $\chi^2$ for $n=4$ parametrization  is slightly smaller than that of $n=1$.  Secondly, as is also shown in figure \ref{fig6},  parametrizations in family II, as a whole, have both smaller areas of the $w(z)$ band and lower minimum value of $\chi^2$ than those in family I,  and among the parametrizations in family II, $n=3$ and $n=4$  are still the two most competitive ones. Finally, among the selected prevalent models, although the area of w(z) band of P2 model becomes relative larger, the relative locations of these models does not changed significantly, and   compared with CPL, UIS, Linear and P2 models, GCG model is still not preferred due to its relative larger $\chi^2_{\rm min}$.  The minor difference between figure \ref{fig6} and figure \ref{fig7} may arise from the data calibration of different data sets among which we shall leave in a future work about a comprehensive comparison.

\begin{table*}
 \centering
 \begin{minipage}{0.8\textwidth}
  \caption{The minima of $\chi^2$ and areas of the $w(z)$ band for different models using 157 Gold SNIa data \citep{gold}.}
  \begin{center}
    \begin{tabular}{c|cccc}\label{table1}
 Model &  $\chi^2_{\rm min}$ &  $w_0$($a_1$ or $\Omega_1$) &  $w_1$($a_2$ or $\Omega_2$) & $s$ \\
\hline
 UIS &  174.365 &  -1.40802$\pm$0.255438 &  1.70941$\pm$0.928001 & 1.69918 \\
 Linear &  174.365 &  -1.39978$\pm$0.249302 &  1.66605$\pm$0.892594 & 2.13868 \\
 P2 &  174.207 &  -4.16234$\pm$2.62176 &  1.67458$\pm$1.06813 & 0.895176 \\
 CPL &  173.928 &  -1.57705$\pm$0.326346 &  3.29426$\pm$1.69727 & 1.62803 \\
 GCG &  177.063 &  0.999827$\pm$0.00663069 &  83.4676$\pm$3209.18 & 1.58943 \\
 Family  I &   &   &   &  \\
 n=1 &  173.928 &  -1.57705$\pm$0.326346 &  3.29426$\pm$1.69727 & 1.62803 \\
 n=2 &  174.606 &  -1.27011$\pm$0.192597 &  6.20395$\pm$3.44687 & 2.13084 \\
 n=3 &  175.09 &  -1.17171$\pm$0.154602 &  12.8437$\pm$7.65729 & 2.71702 \\
 n=4 &  175.444 &  -1.12497$\pm$0.138595 &  26.4417$\pm$16.8124 & 3.37756 \\
 Family-II &   &   &   &  \\
 n=1 &  173.928 &  -1.57705$\pm$0.326346 &  3.29426$\pm$1.69727 & 1.62803 \\
 n=2 &  173.409 &  -1.87262$\pm$0.456452 &  6.62831$\pm$3.29276 & 1.14253 \\
 n=3 &  172.824 &  -2.39635$\pm$0.69199 &  13.7569$\pm$6.65922 & 0.731673 \\
 n=4 &  172.454 &  -3.2745$\pm$1.11672 &  28.3698$\pm$13.8425 & 1.30027 \\
\hline
\end{tabular}
  \end{center}
  \end{minipage}
\end{table*}

\begin{table*}
 \centering
 \begin{minipage}{0.8\textwidth}
  \caption{The minima of $\chi^2$ and areas of the $w(z)$ band for different models using 192 SNIa data \citep{Davis07,Wood07,SNIa}.}
\begin{center}
\begin{tabular}{c|cccc}\label{table2}
Model &  $\chi^2_{\rm min}$ &  $w_0$($a_1$ or $\Omega_1$) &  $w_1$($a_2$ or $\Omega_2$) & $s$ \\
\hline
 UIS &  195.412 &  -1.1192$\pm$0.27732 &  0.0485532$\pm$1.17151 & 2.21778 \\
 Linear &  195.409 &  -1.12628$\pm$0.281052 &  0.0811196$\pm$1.18901 & 2.92336 \\
 P2 &  195.382 &  -0.591863$\pm$1.71977 &  0.166512$\pm$0.679303 & 3.07726 \\
 CPL &  195.411 &  -1.12456$\pm$0.331918 &  0.0961458$\pm$1.89159 & 1.88532 \\
 GCG &  195.529 &  1.00055$\pm$0.00898214 &  95.6168$\pm$1553.63 & 1.62394 \\
 Family  I &   &   &   &  \\
 n=1 &  195.411 &  -1.12456$\pm$0.331918 &  0.0961458$\pm$1.89159 & 1.88532 \\
 n=2 &  195.413 &  -1.11369$\pm$0.21235 &  0.135963$\pm$4.91012 & 3.1127 \\
 n=3 &  195.402 &  -1.12407$\pm$0.181828 &  1.52726$\pm$14.2855 & 5.13255 \\
 n=4 &  195.314 &  -1.15242$\pm$0.164779 &  13.4211$\pm$34.8666 & 7.01332 \\
 Family-II &   &   &   &  \\
 n=1 &  195.411 &  -1.12456$\pm$0.331918 &  0.0961458$\pm$1.89159 & 1.88532 \\
 n=2 &  195.409 &  -1.13475$\pm$0.412811 &  0.203332$\pm$3.09753 & 1.13799 \\
 n=3 &  195.399 &  -1.17258$\pm$0.546306 &  0.631377$\pm$5.28164 & 0.641504 \\
 n=4 &  195.356 &  -1.29367$\pm$0.787516 &  2.28402$\pm$9.61853 & 0.960275 \\
\hline
\end{tabular}
\end{center}
\end{minipage}
\end{table*}

\section{conclusions and discussions}

Traditionally, forming the so-called Bayes factor (likelihood ratio for frequentists) $B_{ij}\equiv L(M_i)/L(M_j)$, where $L(M_{i})$ is called likelihood for the model $M_{i}$ to obtain the data if the model is true, is used in comparison of the cosmological(and/or dark energy) models \citep{baye1,baye2}. Generally, $L(M_i)$ is dependent on the prior probability and the likelihood,which is determined by $\chi^2$, for the model parameters. And when one has no prior to the model parameter, everything is determined by $\chi^2$, which is a measure of the fit to the data. However, according to the above analysis based on the latest SNIa data, there exist lots of cosmological models and parametrizations of dark energy which lead to very similar $\chi_{\rm min}^2$. This is especially true for the results we obtained based on the newly compiled 192 data. The best fitting of any parametrization or model we consider here is within the $1\sigma$ bound of those of  others, see table \ref{table2} or figure \ref{fig7}.  Under this circumstance, how do we compare them? Or what parametrization approach should be used to probe the nature of dark energy in the future experiments? The Bayes approach only works in the condition that fittings of models are distinctly different. When the difference of $\chi^2$ is very small, one should pursue other figures of merit. The above introduced area of $w(z)-z$ band, we think, is such a figure of merit and our point is that both  $\chi^2$ and the area of $w(z)-z$ band should be synthetically  considered for choosing a better parametrization of dark energy in the future experiments.

Bearing this point in mind and according to the results presented in the above section,  we find that the widely used CPL parametrization, which has a very simple interpretation in terms of the scale factor,  is not statistically "special" among the two-parameter parametrization families,  instead $n=3$  in family II, which looks like a variation on the CPL parametrization, is more preferred. Note that CPL parametrization corresponds to the $n=1$ case in both family I and family II and if we also take $n$ as a free parameter, family I and family II are just two three-parameter parametrizations. However, in this work we only consider two-parameter parametrizations and $n$ is not treated as a free parameter. so $n=1$ and $n=3$ actually denote two distinct parametrizations, in which the evolution of $w(z)$ is qualitatively different.

There is an interesting question that whether  the differences among the area of $w(z)$ band are significant enough to single out one parametrization. In our opinion, the answer is somewhat depended on the observational data. As far as the data we used here, the differences among the areas of $w(z)$ band are so significant that we can pick out $n=3$ of the family II as the best parametrization among the models we consider in this work. However, this does \emph{not} mean that the other parametrizations are completely ruled out. For example, the simple CPL parametrization and P2 model still do well to a certain extent.  It should be also pointed out that,  the differences among the areas of $w(z)$ band are much more significant than those among  $\chi_{\rm min}^2$ for both 157 gold data and latest 192 data. This fact indicates that the area of $w(z)$ band is likely to be a good figure of merit, especially in the situation that the value of $\chi^2_{\rm min}$ for different parametrizations are very close.

Generally speaking, the motivation from a physical point of view should be at the top priority when we choose cosmological models. However, it is perfectly clear that in the absence of any compelling dark energy model, the suggested parametrizations are phenomenological. Then the reason why people might prefer a given parametrization is because of its simplicity and also because they feel that it allows us to extract useful information for a very large class of
models, and hopefully the "true" model is one of them. Anyway, to estimate the effects of dark energy one needs to quantify them and parametrization of $w$ has turned out to be an efficient tool in this respect. Therefore, there is a subtle balance between motivation from a physical point of view and fitting results. To help making decisions in this situation, we need to know what is the best achievable fitting result from various models or parametrizations  with the same number of parameters. This will serve as a fiducial criteria for us to choose a \emph{best} model. The figure of merit introduced in this paper is to help to define what is the best.

As is well known, besides SNIa observations,  there exist lots of other experiments probing different aspects of dark energy and  we will have many more data from these experiments \citep{DETF1}.
However, in terms of constraining the evolution of $w(z)$, SNIa approach is the most sensitive and direct one. Other methods, such as CMB and cluster counts, are primarily good for the energy density constraint. But it will be advantageous to test all the parametrization with all the combined data sets in the future. The current analysis in this paper could be  directly generalized to the case with multi-experiments by maximizing the product of the likelihood of each experiment. It is worth noting that the best parametrization of dark energy models for SNIa data may not necessarily be the best one for other observational data. We will report that in a preparing work.

\section*{Acknowledgments}

This work is supported by National Natural Science Foundation of
China under Grant No. 10473007 and No. 10503002 and Shanghai
Commission of Science and technology under Grant No. 06QA14039.

%\bsp

\label{lastpage}

\end{document}